\documentclass[journal]{IEEEtran}

\usepackage{graphicx}
\usepackage{url}
\usepackage{hyperref}
\usepackage{cite}
\usepackage{amsmath, amsthm, amssymb, amsfonts} 
\usepackage{siunitx}
\DeclareMathOperator\sgn{\mathrm{sgn}}
\newcommand{\sumlimits}[1]{\sum_{\makebox[0pt]{$\substack{#1}$}}}

\begin{document}

\title{Delay-Multiply-And-Sum Beamforming for Real-Time In-Air Acoustic Imaging}

\author{
\IEEEauthorblockN{
    Wouter Jansen,
    Walter Daems,
    and Jan Steckel
     }\\
     \IEEEauthorblockA{Cosys-Lab, University of Antwerp, Antwerp, Belgium}\\
    \IEEEauthorblockA{Flanders Make Strategic Research Centre, Lommel, Belgium}\\

\thanks{Corresponding author: Jan Steckel (email: jan.steckel@uantwerpen.be)}
}

\maketitle

\begin{abstract}
In-air acoustic imaging systems demand beamforming techniques that offer high dynamic range and spatial resolution while remaining robust, yet conventional Delay-and-Sum (DAS) beamforming suffers from high sidelobes and low contrast. Conversely, advanced adaptive methods are typically too computationally expensive for real-time field operation. To overcome this trade-off, we propose a higher-order non-linear beamforming method using the Delay-Multiply-and-Sum (DMAS) technique with Coherence Factor weighting, specifically adapted for in-air microphone arrays. By leveraging Newton-Girard binomial expansion, we reduce the complexity of $n$-th order DMAS to $\mathrm{O}(N)$, enabling GPU-accelerated, real-time performance on embedded platforms, with the implementation also made available as open-source. Through validation against simulated and real-world data, we demonstrate that the proposed 5th-order DMAS with CF achieves a dynamic range improvement of nearly $40\text{ dB}$ over DAS, alongside significantly higher image SNR and artifact suppression. We demonstrate real-time viability on various GPU platforms, establishing this approach as a practical, high-performance solution for 3D in-air sonar and industrial acoustic imaging.
\end{abstract}


\section{Introduction}
\label{sec:introduction}
In-air acoustic imaging is a technique in which an array of microphones is used to reconstruct the incoming wave field that is being emitted by one or more sources in the environment. The fundamental goal of this process is to convert the time or frequency-domain signals recorded across an array of microphones into a spatially localized map that visualizes the power distribution. 

This technique has a wide range of applications, one of which is the detection of pressurized gas or air leaks in industrial environments \cite{guentherAutomatedDetectionCompressed2016, steckelUltrasoundbasedAirLeak2014}. High-pressure gas escaping through a small orifice generates acoustic energy due to the turbulent, high-velocity flow. The ability to precisely localize these leaks in complex piping systems is vital for energy efficiency, environmental compliance, and worker safety. Furthermore, non-contact fault detection and condition monitoring of rotating machinery is another major driver \cite{alousifMachineryFaultDetection2021, dadoucheSensitivityAirCoupledUltrasound2008, steckelToolWearPrediction2024, verellenBeamformingAppliedUltrasound2021}. Mechanical faults, such as bearing wear produce ultrasonic emissions long before they generate significant audible noise or heat, making acoustic imaging a viable tool for predictive maintenance. It is also used in aeroacoustics for the analysis of wing performance \cite{merino-martinezReviewAcousticImaging2019}. Finally, ultrasonic imaging is of importance for in-air sonar sensing\cite{allevatoEmbeddedAircoupledUltrasonic2020, jansenStabilizedAdaptiveSteering2024, moto:c:irua:166448_kers_comp}. It is used as a pulse-echo sensor in, for example, (3D) robotic navigation and mapping. It allows for robust depth sensing and can be of critical value for autonomous systems operating in visually obscured environments (by dust, rain, fog, etc.) where other sensors such as cameras may fail.

\begin{figure*}
    \centering
    \includegraphics[width=1\linewidth]{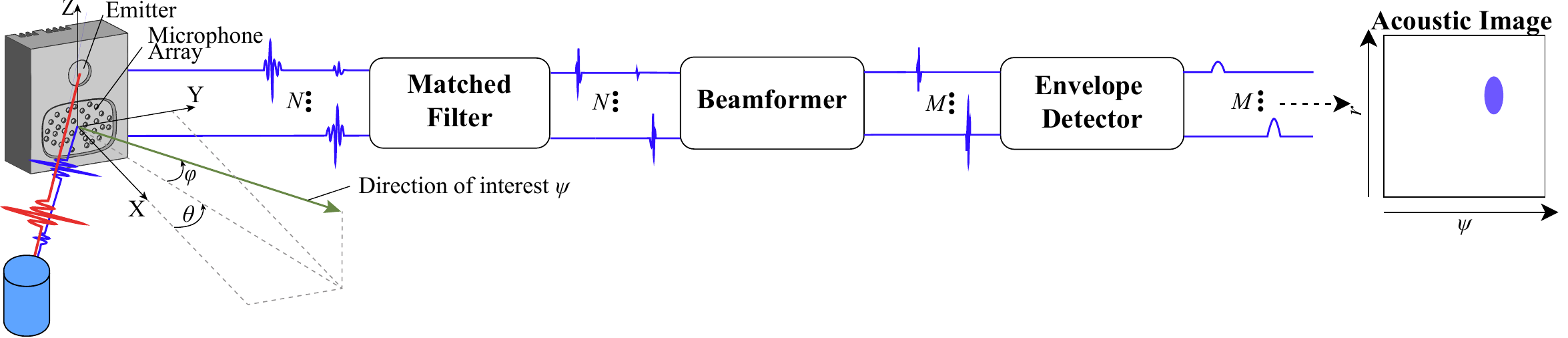}
    \caption{On the left side of the figure we depict a 3D in-air sonar device (eRTIS\cite{laurijssen2025ruggedizedultrasoundsensingharsh}) as used in the experiments with the typical situation of an active pulse-echo measurement. The signal from the emitter reflects on an object and is received by all microphones, which then goes through the several signal processing steps to result in the final acoustic image. The steps used in this processing pipeline are matched filtering (pulse compression), beamforming and a final envelope extraction phase. In this paper, we will focus on improvements of the beamforming stage, which converts the signals of the microphone array into a spatialized set of signals. The figure furthermore shows the coordinate system with the azimuth angle $\theta$ and elevation angle $\varphi$, used to define directions of interest $\psi$ or sound locations. In the right sketch of the acoustic image, the vertical axis uses the range dimension $r$, converted from the time domain $t$.}
    \label{fig:ertis_and_processing}
\end{figure*}

A general recurrence in these applications that make use of acoustic imaging techniques is the usage of spatial signal processing. This implies the usage of some kind of sensor array, and an algorithm that converts the multitude of input signals into some kind of spatially localized signal. Array signal processing has a very long and rich history, and several seminal books and papers have been written on this subject \cite{treesOptimumArrayProcessing2002,krimTwoDecadesArray1996,pesaventoThreeMoreDecades2023}. When taking a zoomed-out view, the domain can be simplified and categorized into the following categories: 

\begin{itemize}
    \item Spatial filtering: Delay-and-Sum (DAS) \cite{krimTwoDecadesArray1996,van1988beamforming}
    \item Super-resolution filtering: Minimum Variance Distortionless Response (MVDR) \cite{treesOptimumArrayProcessing2002,moto:c:irua:172542_vere_ultr}
    \item Spatial spectrum estimator: Multiple Signal Classification (MUSIC) \cite{moto:c:irua:166448_kers_comp,labyed2013super}
    \item Compressed sensing: Sparse representations \cite{steckelSparseDecompositionInair2014,sonIrregularMicrophoneArray2022}
    \item Machine-learning based methods \cite{steckel2024echopt2,lafontaine2023denoising,mansour2018sparse}
\end{itemize}
The core differences between these listed techniques lie in how they approach the fundamental trade-off between image quality characteristics such as spatial resolution, sidelobe level, dynamic range, and robustness to noise as well as to correlated sources. Furthermore, computational cost has to be taken into account for practical, real-world systems. While advanced methods (like MVDR and MUSIC) offer superior resolution compared to conventional DAS, they often incur high computational complexity, and rely on assumptions that can degrade real-time performance and dynamic range in complex, real-world ultrasonic environments. More specifically, most of the so-called super-resolution techniques use spatial statistics of the incoming wavefield, requiring multiple measurements (snapshots) for the estimation thereof. In acoustic imaging, the luxury of multiple snapshots is not available, as this would incur a measurement time that is too large given the slow speed of sound\cite{moto:c:irua:172542_vere_ultr}.

Given the constraints of real-time, high dynamic range acoustic imaging, there is a need for a spatial signal processing algorithm that offers better image quality than for example DAS, without incurring the computational complexity and model dependence of MVDR or MUSIC. In this paper, we take inspiration from medical ultrasonic imaging, a field where contrast and resolution are paramount, from their usage of non-linear beamforming methods, which have been shown to drastically enhance image contrast and suppress sidelobe levels. In the biomedical field non-linear beamforming has been extensively explored to enhance contrast and resolution. We focused on the application of the Delay-Multiply-and-Sum (DMAS) \cite{lim2008confocal} combined with the optional Coherence Factor (CF) \cite{coherencehollman1999} weighting for in-air acoustic imaging. This methodology has demonstrated superior contrast ratios and increased range resolution in various biomedical applications, ranging from standard piezoelectric transducers to recent PMUT-based systems \cite{Mantalenadmasius2025}, but remains relatively unexplored in the context of in-air acoustic arrays.

The core principle of DMAS is to introduce a non-linear multiplication of delayed (or so-called pre-steered or pre-aligned) microphone signals.
The DMAS algorithm as implemented for biomedical ultrasound by G. Matrone et al. \cite{dmasmatrone} has seen various adaptations to reduce its computational load, such as the algebraic factorization proposed by A. Ramalli et al. \cite{Ramalli2017}, baseband implementations \cite{shen2021}, and double-stage approximations \cite{Mozaffarzadeh2018}. While these optimization strategies have proven effective for standard second-order non-linearities, and specific higher-order formulations have been investigated for photoacoustic imaging \cite{mulani2022higher}, the specific requirements of in-air sonar necessitate an efficient, generalized approach capable of handling arbitrary orders.

The primary contributions of this paper are twofold:
\begin{enumerate}
\item We propose and detail the implementation of the DMAS-CF technique specifically for ultrasonic in-air microphone arrays, addressing its unique challenges. We propose an efficient implementation method based on the Newton-Girard binomial expansion theorem and accelerated implementation on GPU hardware.
\item We provide a direct, quantitative comparison between the DMAS-CF method and others in both simulation and real-world experiments, demonstrating significant improvements of the DMAS approaches over classical beamforming.
\end{enumerate}
This paper is organized as follows: Section \ref{sec:methods} reviews the theoretical basis and implementation of our proposed solution. Section \ref{sec:results} details both quantitative comparison in simulation and shows real-world experimental results validating its real-time usage. Finally, section \ref{sec:conclusion} offers conclusions and directions for future work.

\begin{figure*}[ht!]
    \centering
    \includegraphics[width=1\linewidth]{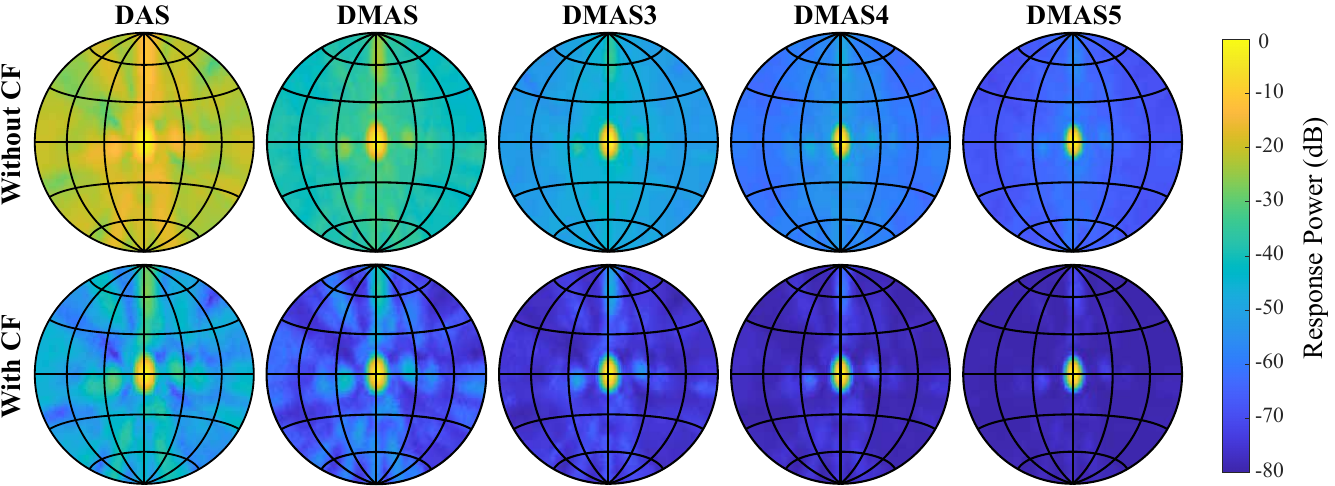}
    \caption{Directional component of the point-spread function in both azimuth and elevation, plotted using a Lambert Equal Area Projection. The resulting PSF shows a higher dynamic range for increasing orders of DMAS processing. Furthermore, the addition of the CF post-processing step further increases the dynamic range of the PSF. Gridlines are spaced by \SI{30}{\degree}.}
    \label{fig:PSF_results_direction}
\end{figure*}

\section{Methods}
\label{sec:methods}
This section details the mathematical and computational methodology used to generate an acoustic image and perform source localization from raw microphone signals. The contribution of this paper focuses on the beamforming step of the imaging process, which will be reflected in this section as well. However, the overall process, as implemented in our general purpose signal processing pipeline\cite{jansen2019gpu}, follows a sequential order with additional optional pre- and post-processing steps and filters, also visualized in Fig. \ref{fig:ertis_and_processing}. 

\noindent These steps are: 
\begin{enumerate}
    \item \textbf{Matched Filtering:} The raw signal is optionally first convolved with the known emitted source signal to produce a matched-filtered signal enhancing the signal-to-noise ratio and increasing the range resolution (sometimes referred to as pulse compression in the literature).
    \item \textbf{Beamforming:} The core beamforming algorithm is applied to compute the acoustic image.
    \item \textbf{Envelope Detection:} The absolute value of the signal is taken and then low-pass filtered to extract its envelope.
\end{enumerate}
The system consists of an array of $N$ elements, in our case microphones. The received signal at the $i$-th microphone is denoted as $m_i(t)$. To form an image, a set of $M$ directions of interest is defined, which effectively constitutes the spatial grid of pixels or voxels where the acoustic intensity will be estimated. For each direction $\psi$, a time delay $\tau_{i,\psi}$ is calculated for each $i$-th microphone, representing the propagation time of an acoustic wave from the direction to the $i$-th microphone. In the case of 3D beamforming, these directions are defined by their azimuth and elevation angles $(\theta,\varphi)$. Fig. \ref{fig:ertis_and_processing} visualizes the coordinate system and the signal processing pipeline. 

The delays for these directions are pre-computed and stored in a delay matrix look-up table. The pre-steering operation is to time-shift the received signals according to these delays, effectively compensating them. This temporal alignment ensures that signal components originating from the target direction are in phase across all channels, a prerequisite for coherent summation. For a given time $t$ and direction $\psi$, we define the delayed signal $x_i(t,\psi)$ as:
\begin{equation}
    x_i(t,\psi) = m_i(t + \tau_{i,\psi})
    \label{eq:signal_delay}
\end{equation}
For simplicity, in the following equations, we will omit the $(t,\psi)$ arguments and use $x_i$ to denote the appropriately pre-steered signal for the specific output direction being computed. 

\subsection{Delay-and-Sum (DAS) Beamforming}
\label{sec:das}
Delay-and-Sum or Bartlett beamforming is the oldest and most direct beamforming method, as this method computes the output by coherently summing the delayed signals from all $N$ microphones:
\begin{equation}
    S_{\textsl{DAS}} = \sum_{i=1}^{N} x_i
    \label{eq:das}
\end{equation}
DAS, as a non-adaptive beamformer, treats desired signals and noise/clutter identically, leading to high sidelobes and low image resolution and contrast, which is especially problematic in the noisy, attenuating environment of in-air acoustic imaging.

\subsection{Delay-Multiply-and-Sum (DMAS) Beamforming}
\label{sec:dmas}
DMAS is a non-linear technique that involves multiplying combinations of delayed signals before summation \cite{lim2008confocal, dmasmatrone}. It is generally described by:
\begin{equation}
    S_{\textsl{DMAS}} = \sum_{i=1}^{N-1}\sum_{j=i+1}^{N} \sgn(x_ix_j) \cdot \sqrt{|x_ix_j|}
    \label{eq:dmas}
\end{equation}
The multiplication will cause lower sidelobes and a higher dynamic range by acting as a correlation function, which in effect suppresses low-coherence parts in the image signals. As these low-coherence parts are typically sidelobes in the image, and not real reflected signals, this operation increases image quality. High-order DMAS methods have been proposed to further suppress more coherent artifacts \cite{mulani2022higher}. Let the $n$-th signed root of the delayed signal $x_i$ be:
\begin{equation}
    s_i^{(n)} = \sgn(x_i) \cdot \sqrt[n]{|x_i|}
\end{equation}
Using this $n$-th root signal $s_i^{(n)}$, the higher-order version of DMAS (order $n$) is given by: 
\begin{equation}
  S_{\textsl{DMAS}}^{(n)} = \sum_{I \in \binom{1:N}{n}} \prod_{i \in I} s_{i}^{(n)}
  \label{eq:dmasn}
\end{equation}
with $\binom{1:N}{n}$ the set of all possible combinations of $n$ indices out of the set of integer numbers from $1$ to $N$.
$S_{\textsl{DMAS}}^{(n)}$ is the sum over all unordered $n$‑tuples of distinct microphones of the product of their signed $n$‑th roots. This naive implementation of the $n$-th order DMAS has a complexity of $\mathrm{O}(N^n)$ per output pixel of the image (i.e. for each $(t,\psi)$), which for real-time operation would be too high while also trying to achieve sufficient image resolution by having enough directions of interest. 

\subsection{Efficient Higher-Order DMAS Beamforming}
\label{sec:dmasn}
The $n$-th order DMAS output is by definition the $n$-th elementary symmetric polynomial, $E_n$, of the $N$ signed root signals $\{s_1^{(n)}, \ldots, s_N^{(n)}\}$:
\begin{equation}
    S_{\textsl{DMAS}}^{(n)} = E_n(s_1^{(n)}, \ldots, s_N^{(n)})
    \label{eq:dmas_as_En}
\end{equation}

\begin{figure*}[!ht]
    \centering
    \includegraphics[width=1\linewidth]{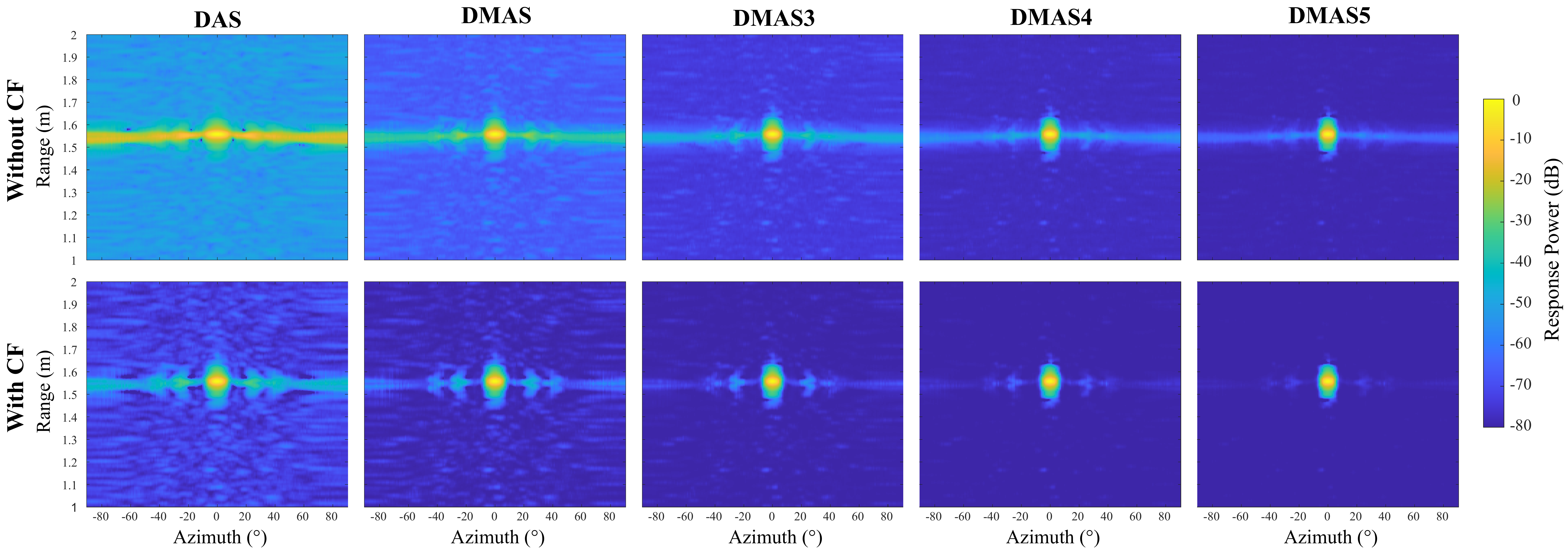}
    \caption{Horizontal slice of the Point Spread Function (PSF) of the acoustic imaging system, shown in a range/azimuth plot (for elevation equal to 0 degrees). Similar to the directional slice through the PSF, an increase in dynamic range can be observed for increasing orders of DMAS processing, with an additional gain to be made when applying the CF post-processing step. No discernible increase in the range resolution can be observed, as is expected from the theory underpinning the far-field imaging system.}
    \label{fig:psf_range}
\end{figure*}

As stated before, directly evaluating this from its combinatorial definition as in Equation \eqref{eq:dmasn} is computationally infeasible for real-time applications.

To achieve an efficient implementation, we use the Newton-Girard identities \cite{macdonald1998symmetric}, which can provide an explicit formula for $E_n$ in terms of the power sums of its arguments. The first step is to compute the necessary power sums, $P_k^{(n)}$, for $k=1, \ldots, n$:
\begin{equation}
P_k^{(n)} = \sum_{i=1}^{N} (s_i^{(n)})^k = \sum_{i=1}^{N} \left( \sgn(x_i) \cdot \sqrt[n]{|x_i|} \right)^k
\label{eq:power_sum}
\end{equation}
To keep the notation simple the dependency on $s_i^{(n)}$ has been left out of the notation for $P_k^{(n)}$. The general, non-recursive relationship between $E_n$ and the power sums $P_k^{(n)}$ can be formulated by summing over all integer partitions of $n$:
\begin{equation}
E_n = \sumlimits{k_1, k_2, \ldots, k_n \ge 0 \\ k_1 + 2k_2 + \cdots + nk_n = n} (-1)^{n - \sum_{i=1}^n k_i} \cdot \prod_{i=1}^{(n)} \frac{(P_i^{(n)})^{k_i}}{k_i! \cdot i^{k_i}}
    \label{eq:girard_newton_general}
\end{equation}
While this general formula appears complex, it can be pre-expanded for a fixed order $n$ to yield a highly optimized, explicit equation, ideal for implementing in optimized non-recursive signal processing algorithms. For the orders relevant to this work (e.g., $n=2$ to $5$), these expansions are:
\begin{equation}
    S_{\textsl{DMAS}}^{(2)} = S_{\textsl{DMAS}} = \frac{1}{2} \left( (P_1^{(2)})^2 - P_2^{(2)} \right)
    \label{eq:dmas2_impl}
\end{equation}
\begin{equation}
    S_{\textsl{DMAS}}^{(3)} = \frac{1}{6} \left( (P_1^{(3)})^3 + 2P_3^{(3)} - 3P_1^{(3)}P_2^{(3)} \right)
    \label{eq:dmas3_impl}
\end{equation}
\begin{equation}
\begin{split}
    S_{\textsl{DMAS}}^{(4)} = \frac{1}{24} \Big( & (P_1^{(4)})^4 - 6P_4^{(4)} + 3(P_2^{(4)})^2 \\
    & - 6P_2^{(4)}(P_1^{(4)})^2 + 8P_3^{(4)}P_1^{(4)} \Big)
\end{split}
\label{eq:dmas4_impl}
\end{equation}
\begin{equation}
\begin{split}
    S_{\textsl{DMAS}}^{(5)} = \frac{1}{120} \Big( & (P_1^{(5)})^5 - 10P_2^{(5)}(P_1^{(5)})^3  + 15(P_2^{(5)})^2 P_1^{(5)} \\ 
    & + 20P_3^{(5)}(P_1^{(5)})^2 - 20P_3^{(5)}P_2^{(5)} \\
    & - 30P_1^{(5)}P_4^{(5)} + 24P_5^{(5)} \Big)
\end{split}
\label{eq:dmas5_impl}
\end{equation}
Note that order 2 is equivalent to DMAS as in Equation \eqref{eq:dmas}. 
Furthermore, the computational efficiency of our Newton-Girard implementation for the second order is mathematically equivalent to the factorization-based F-DMAS approach established by A. Ramalli et al \cite{Ramalli2017}. However, unlike prior factorization or multi-stage methods \cite{Mozaffarzadeh2018} which are often derived specifically for quadratic or cubic non-linearities, the Newton-Girard identities provide a generalized, exact, and non-recursive formulation for any arbitrary order $n$. For any given order $n$, our method involves first computing the power sums $P_1^{(n)}, \ldots, P_n^{(n)}$ in $\mathrm{O}(N)$ time, and then substituting these values into the corresponding explicit formula above to obtain $S_{\textsl{DMAS}}^{(n)} = E_n$. This non-recursive, explicit method is highly efficient for a fixed set of orders.

\subsection{Coherence Factor (CF) Weighting}
\label{sec:cf}
The Coherence Factor (CF) is an optional spatial weighting applied to the beamformer output to suppress incoherent noise and sidelobes \cite{coherencehollman1999}. It measures the spatial coherence of the delayed signals across the array aperture, and is defined as the ratio of the coherent power to the incoherent power.  The specific formulation we implemented can be written as follows:
\begin{equation}
    \textsl{CF} = \frac{\left(\sum_{i=1}^{N} x_i\right)^2}{N \cdot \sum_{i=1}^{N} x_i^2}
    \label{eq:cf}
\end{equation}

A small, positive number is typically added to the denominator to prevent division by zero. The final output is the product of the beamformed signal and the CF:
\begin{equation}
    S^{(n)}_{\textsl{DMAS,CF}}(t,\psi) = S^{(n)}_{\textsl{DMAS}}(t,\psi) \cdot \textsl{CF}(t,\psi)
\end{equation}
In this final formula, we added the time and direction dependency $(t,\psi)$ explicitly to remind the reader of these dependencies. Note that the coherence factor is independent of the spatial beamforming technique chosen and can therefore also be applied to DAS and other techniques.

\begin{figure*}[!ht]
    \centering
    \includegraphics[width=1\linewidth]{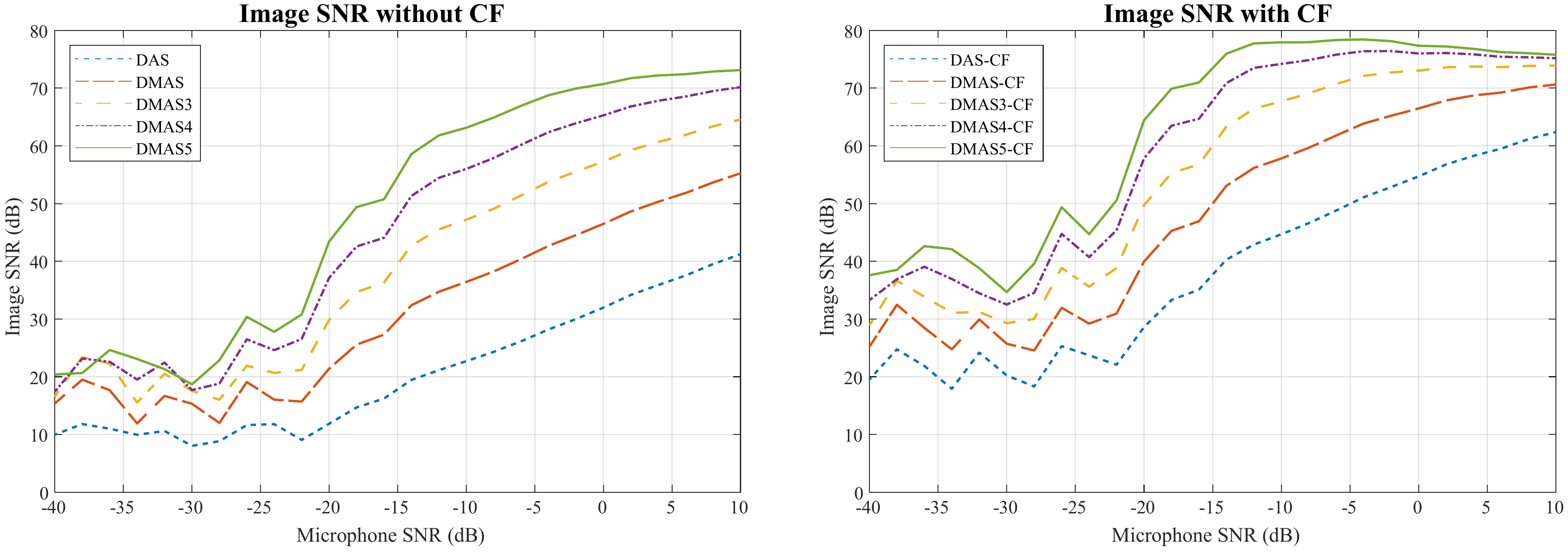}
    \caption{Image SNR results for varying input noise levels. The image SNR quantities were calculated for input microphone noise levels $\mathrm{mic}$ ranging from -\SI{40}{\dB} to \SI{10}{\dB}. The left panel shows the SNR results obtained without CF post-processing, while the right panel shows the results with CF post-processing applied. The plots illustrate the superior noise suppression of the DMAS-class algorithms, as evidenced by increasing image SNR values with higher DMAS orders. The inclusion of the CF post-processing step further enhances the image SNR across all algorithms.}
    \label{fig:image_snr}
\end{figure*}

\subsection{Computational GPU Acceleration and Code Availability}
\label{sec:gpu}
Even with the proposed efficient implementation, the entire signal processing is computationally demanding, especially for real-time processing on embedded or edge-computing devices. To achieve high performance, we have implemented the necessary beamforming operations on GPU hardware. We implemented a kernel function that achieves parallelism by launching a separate thread for each output pixel (i.e., for each $(t,\psi)$). The efficient $\mathrm{O}(N)$ DMAS formulation is critical, as it allows each thread to compute its output by iterating only once over the $N$ microphones. Our accelerated implementation is written in NVIDIA's CUDA framework, and the source code can be found on our GitHub page: \textit{https://github.com/Cosys-Lab/Acoustic-Imaging-Toolbox}. . This repository is structured as a Matlab Toolbox containing the proposed DMAS-CF methods implemented in native Matlab code, CPU-accelerated C, and GPU-accelerated CUDA.

\section{Experimental Results And Validation}
\label{sec:results}
To evaluate the performance and merit of the DMAS algorithms for in-air sensing, we conducted several experiments. We want to note that in our results, we restrict our comparison solely to DAS against our proposed method because we want to focus on real-world conditions for a broad field of realistic application scenarios. This means we want to operate using only single snapshot, precluding the use of adaptive methods like MVDR which require multiple snapshots to accurately estimate data statistics. Furthermore, we require a method that functions effectively with any arbitrary microphone array configuration. Indeed, single-snapshot adaptations for super-resolution techniques exist, but require the application of spatial smoothing, which puts constraints on the array layout\cite{moto:c:irua:172542_vere_ultr}. Finally, the imaging systems we want to apply these techniques to are inherently broadband, with typical frequency ranges from \SI{25}{\kHz} to \SI{50}{\kHz}. 

For these wideband applications, techniques like DAS and DMAS inherently function well as they are time-domain processing algorithms, unlike some frequency-domain or narrowband adaptive techniques such as MVDR or MUSIC.

\subsection{Simulation Analysis}
\label{sec:results_sim}
In order to quantify the performance of  DMAS-class of beamformers in in-air sonar applications, we did several simulations of key metrics of this imaging system. As a first validation, we calculated the Point-Spread Function (PSF) of the whole imaging system. A PSF is a commonly used characterization of an imaging system, and describes the response of the system to a Dirac point source. In an active imaging system the point-spread function consists of three dimensions: two direction dimensions and one range dimension. 

In Fig. \ref{fig:PSF_results_direction} we have plotted the directional part of the PSF in the noise-free case. From this figure it becomes apparent that the dynamic range of the point-spread function becomes much greater for increasing DMAS orders, reaching almost \SI{80}{\dB} in the DMAS5 class. Compared to the approximated dynamic range of \SI{40}{\dB} of DAS, the improvement is substantial. Furthermore, the CF post-processing step further increases the dynamic range of the PSF. Little effect can be observed on the width of the main-lobe of the PSF. While the results in Fig. \ref{fig:PSF_results_direction} show significant contrast enhancement, it should be noted that non-linear beamformers achieve this largely through dynamic range alteration rather than an increase in fundamental information content \cite{rindal2019}. These methods non-linearly map the signal, which suppresses sidelobes but may not necessarily improve the underlying detectability of targets in all noise conditions. However, for in-air applications where visual localization and automated thresholding are key, this enhanced contrast remains a significant practical advantage.
\newpage
A second view of the point-spread function can be made in the range/direction plane, more specifically in the horizontal plane for an elevation of \SI{0}{\degree}. Effectively, the PSF is sliced in the horizontal plane, and shown in a range/azimuth plot. The resulting ensemble of these PSFs can be found in Fig. \ref{fig:psf_range}. The dynamic range of the PSF improves for increasing orders of DMAS processing, with an additional gain in the dynamic range when applying the CF post-processing step. As expected, the range resolution is not impacted by the DMAS algorithms, because of the separability of the range and direction dimensions in far-field broadband imaging systems, highlighting the function of DMAS as a pure beamforming operation. 
\begin{figure*}[!ht]
    \centering
    \includegraphics[width=1\linewidth]{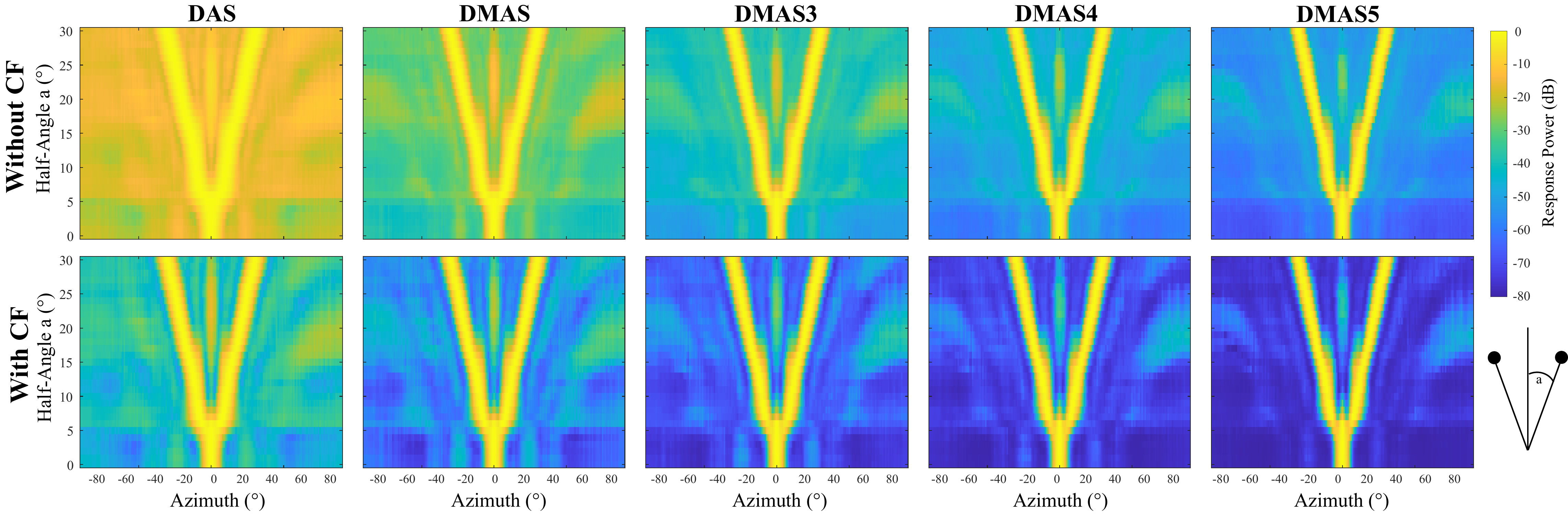}
    \caption{Spatial resolution analysis of DMAS-class beamformers. Two reflectors were simulated at a fixed range of \SI{1.5}{\m}, symmetrically positioned around the X-axis with a varying half-angle $a$. The acoustic image $I$ was computed in the horizontal plane, and the angular responses at the reflector range were aggregated over all inner-angle configurations. The results show that while DMAS methods improve the dynamic range, the spatial resolution remains largely unchanged, confirming that it is primarily determined by the microphone array geometry rather than the beamforming algorithm.}
    \label{fig:innerangle}
\end{figure*}
The literature of medical imaging does highlight increased range resolution when applying DMAS, which can be explained by the fact that in medical imaging, the imaging takes place in the near-field regime (with the sensor array not being considered "small" in respect to the imaging scene). Therefore, range-focusing through beamforming is possible, which causes an effect on the range resolution of the DMAS operations. As with in-air sonar sensing we typically operate in a pure far field regime (with array apertures of \SI{10}{\cm} diameter, and imaging ranges spanning up from \SI{5}{m} to \SI{10}{\m}), the separability of the range and directional imaging steps holds, and therefore, the lack of increase in range resolution when applying DMAS is expected.

In a second experiment, we calculated the signal-to-noise ratio (SNR) of the acoustic image for a varying SNR of the input microphone signals, using the following procedure: Let the amplitude of the echo signal for a source placed at broadside of the array at \SI{1}{\m} range in the microphone channels $m_i(t)$ be equal to 1. We add a white gaussian noise signal $n_i(t)$ to the microphone signals in the following way:
\begin{equation}
    m_i^n(t) = m_i(t) + \eta \cdot n_i(t)
\end{equation}
with a scaling factor $\eta$ equal to:
\begin{equation}
    \eta = 10^{-\frac{1}{20} \mathrm{SNR}_{\mathrm{mic}}}
\end{equation}
We then calculated the acoustic images $I$ in the horizontal plane. The resulting image is normalized such that its maximum value equals one, after which we compute the average off-target energy in the image:
\begin{equation}
E_{\text{off}} = \frac{1}{N_{\text{off}}} \sum_{(r,\theta) \in \Omega_{\text{off}}} I(r,\theta)
\end{equation}
where $\Omega_{\text{off}}$ denotes the set of off-target pixels and $N_{\text{off}}$ is their total number. We then use this quantity $E_{\text{off}}$ as a proximal value for the image SNR:
\begin{equation}
\mathrm{SNR}_{\text{image}} = 20 \log_{10} \left( \frac{1}{E_{\text{off}}} \right)
\end{equation}

We calculated these image SNR quantities for an input range of $\mathrm{mic}$ ranging from -\SI{40}{\dB} to \SI{10}{\dB}. The results of these simulations can be found in Fig. \ref{fig:image_snr}, with the panels showing the SNR results for the imaging with and without coherence factor post-processing on the left and right respectively. From these plots the superior noise suppression of the DMAS-class of algorithms can be observed, with increasing image SNR values for increasing DMAS orders. The addition of the CF post-processing step further increases the image SNR over all algorithms. In Table \ref{tab:snr_cf_gain_0db} we have highlighted the results at a reference microphone SNR of \SI{0}{\dB} to further quantify the improvements in SNR by DMAS-CF. 

\begin{table}
\caption{Impact of CF and Beamforming Order on Image SNR at \SI{0}{\dB} Input}
\label{tab:snr_cf_gain_0db}
\resizebox{\columnwidth}{!}{%
\begin{tabular}{l|r|r|r|r|r}
\textbf{Algorithm} & \multicolumn{2}{c|}{\textbf{No CF}} & \multicolumn{2}{c|}{\textbf{With CF}} & \textbf{CF Gain} \\ \cline{2-5}
 & \textbf{SNR (dB)} & \textbf{+$\Delta$ (dB)} & \textbf{SNR (dB)} & \textbf{+$\Delta$ (dB)} & \textbf{+$\Delta$ (dB)} \\ \hline
\textbf{DAS}   & 32 & --     & 55 & --     & +23 \\
\textbf{DMAS}  & 46 & 14 & 66 & 11 & +20 \\
\textbf{DMAS3} & 58 & 26 & 73 & 18 & +15 \\
\textbf{DMAS4} & 65 & 33 & 76 & 21 & +11 \\
\textbf{DMAS5} & 71 & 39 & 78 & 23 & +7
\end{tabular}%
}
\end{table}

As a final quantification experiment, we measured the spatial resolution of the DMAS-class beamformers. For this, we implemented a simulation in which two reflectors were spaced at the same range (\SI{1.5}{\m}) and were placed symmetrically with respect to the X-axis with a half-angle $a$, as can be seen in Fig. \ref{fig:innerangle} on the bottom right corner. We then calculated the acoustic image $I$ in the horizontal plane, and selected the range slice corresponding to the reflector range. We then aggregated the angular responses for all inner-angle combinations of reflectors in the scene, which are the results shown in Fig. \ref{fig:innerangle}. From these plots it becomes clear that the main benefit of the DMAS methods is an increase in dynamic range, but not in spatial resolution, which remains mostly constant over the spacing of the targets. This confirms the hypothesis that the spatial resolution mainly originates from the layout of the microphone array, and cannot be drastically improved by applying the DMAS-class algorithms for beamforming.

\begin{figure*}[!ht]
    \centering
    \includegraphics[width=0.98\linewidth]{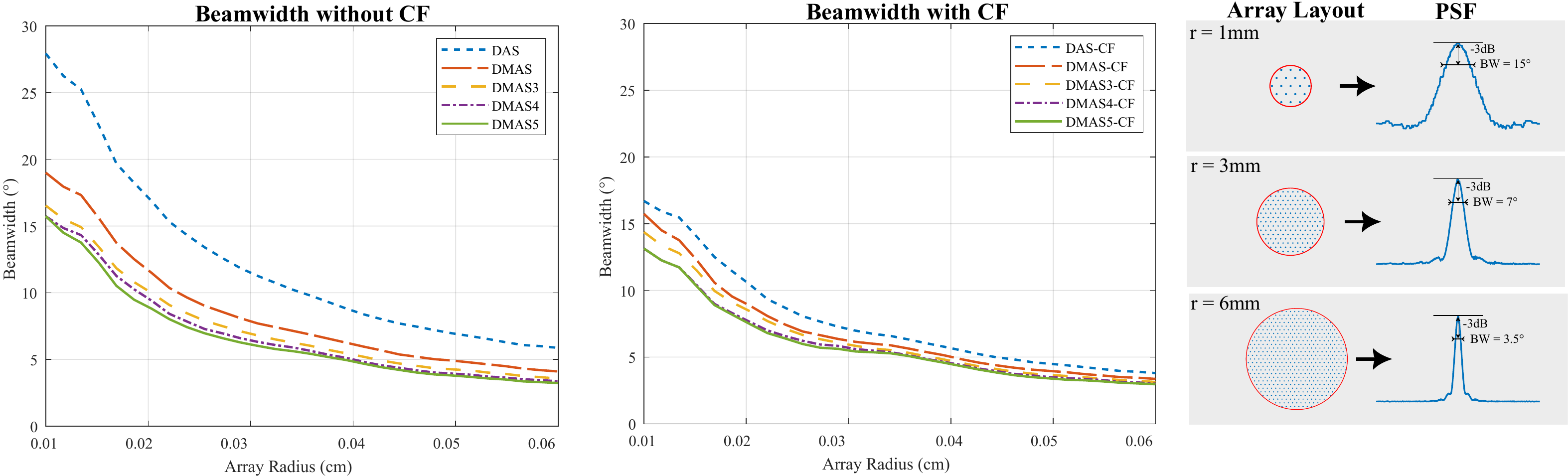}
    \caption{Beamwidth analysis of DMAS beamformers for varying array sizes. The left panel shows the -\SI{3}{\dB} beamwidth of the point-spread function (PSF) as a function of array radius for different beamformer implementations. Circular arrays were simulated with uniformly spaced microphones arranged on a hexagonal lattice with a \SI{5}{\mm} edge length, resulting in array radii between \SI{1}{\cm} and \SI{6}{\cm} (19--513 microphones). The right panel illustrates representative PSFs for small and large arrays, demonstrating the expected narrowing with increasing array size. The results indicate that the dominant reduction in beamwidth stems from the array geometry rather than the beamformer order, with all implementations converging to similar spatial resolutions for larger, yet still practical, array dimensions.}
    \label{fig:psf_vs_arraysize}
\end{figure*}

To support the claim that the spatial resolution of the imaging system only marginally depends on the order of DMAS, we calculated the beamwidth of the various beamformer implementations for varying microphone array sizes. We simulated circular arrays with uniformly spaced microphones inside of the array, and varied the radius of the arrays from \SI{1}{\cm} to \SI{6}{\cm}, for which the microphone count varied from 19 to 513 microphones (as we kept the spacing between the microphones constant on an hexagonal lattice with an edge length of \SI{5}{\mm}). For each of the array shapes we then calculated the point-spread function in the horizontal plane and calculated the -\SI{3}{\dB} beamwidth of the PSF. This does not cause a loss of generality, as the PSF of a circularly symmetric array is circularly symmetric as well. We performed this calculation for all beamformer implementations, and plotted the resulting beamwidths as a function of the array radius in Fig. \ref{fig:psf_vs_arraysize}. The right panel details the experiment and illustrated the more narrow PSF as the array size increases. Both graphs show that the major decrease in beamwidth is caused by the array shape, and less so for the various beamformer implementations, with the beamwidths converging for the larger (yet still very practical) array sizes. This convergence of spatial resolution for larger arrays coincides with the findings depicted in Fig. \ref{fig:innerangle}, where it could be observed that the spatial resolution is not greatly increased using the DMAS-class of beamformers. 

\subsection{Real-World Performance}
\label{sec:results_realworld}
\begin{figure*}
    \centering
    \includegraphics[width=1\linewidth]{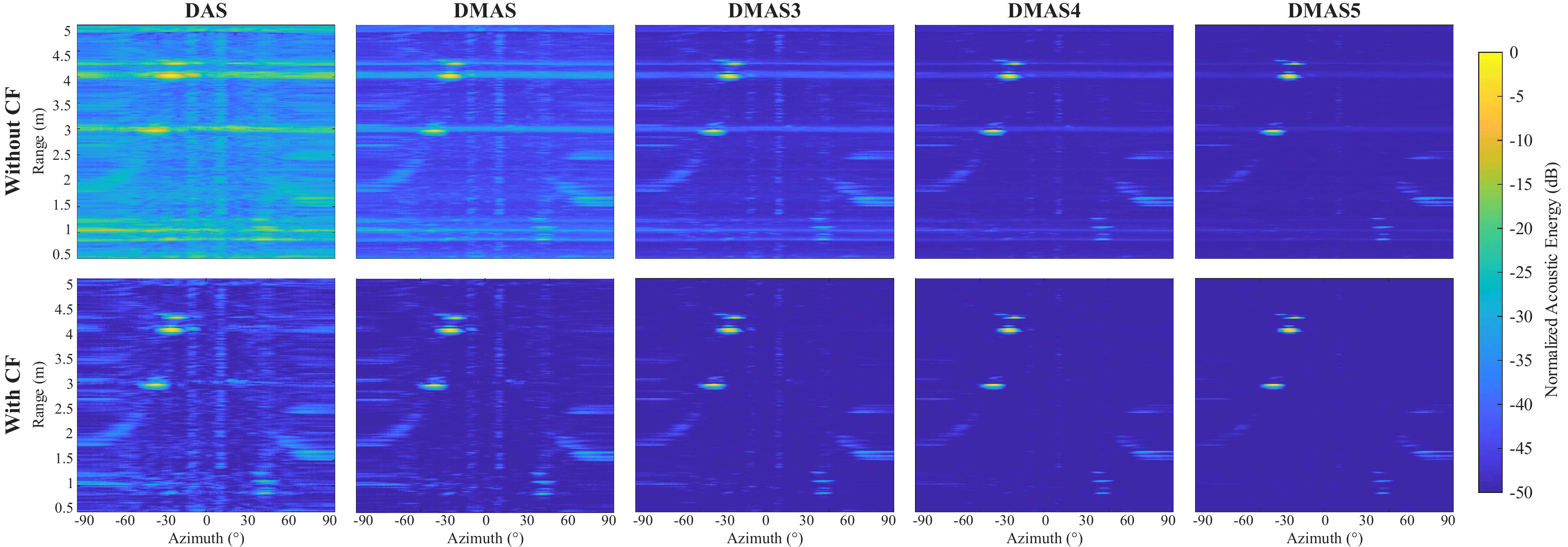}
    \caption{A single acoustic image frame from an indoor scene calculated using the different algorithms proposed in this paper. The increased dynamic range and noise suppression of the higher-order DMAS implementations becomes apparent from these results. In addition, the CF post-processing step further increases the image quality and clarity.}
    \label{fig:single_real_frame}
\end{figure*}

A dataset was recorded using an in-air 3D sonar called eRTIS\cite{laurijssen2025ruggedizedultrasoundsensingharsh}, mounted on a mobile robot driving around. It features a 32-element MEMS microphone array and a broadband capacitive transducer, emitting a \SI{2.5}{\ms} broadband chirp call between \SI{25}{\kHz} and \SI{50}{\kHz}. These measurements were then used to create acoustic images with all beamforming methods, from DAS to DMAS5, and with and without the optional coherence factor post-processing. Beamforming was performed in 3601 directions with their azimuth angle ranging between -\SI{90}{\degree} to \SI{90}{\degree}, with an angular spacing of \SI{0.05}{\degree} to create a horizontal scan image at elevation angle  \SI{0}{\degree}. The recorded signals had 163 840 samples, with a duration of \SI{0.0364}{\s} at \SI{450}{\kHz}. All images were constructed using the same matched filter and envelope detection using a low-pass filter at 5kHz\cite{jansen2019gpu}. A single frame of this dataset is shown in Fig. \ref{fig:single_real_frame}, and a full video of the dataset with all frames processed is available as supplementary material and online\cite{dmascosysyoutube2025}. 

From these images, the increased dynamic range of the DMAS techniques can be observed when compared to the naive DAS implementation as well as the coherence-factor weighing increasing the contrast of the image even further as expected from the literature.

\begin{figure*}
    \centering
    \includegraphics[width=1\linewidth]{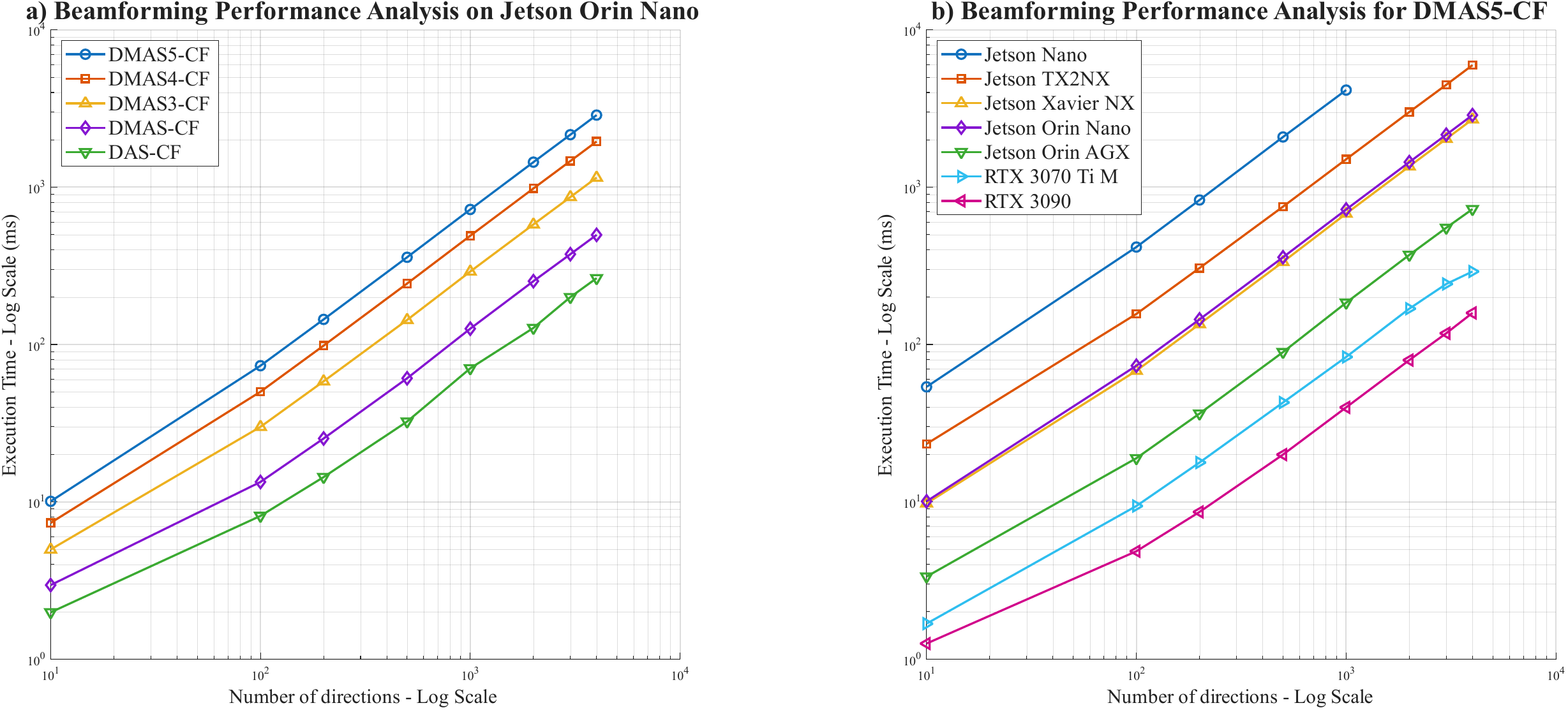}
    \caption{A visualization of the average execution times analysis with (a) a comparison on one mid-range GPU platform of all variants of the proposed beamforming algorithm, and (b) a comparison of NVIDIA GPU platforms tested for DMAS5-CF. Note that the Jetson Nano platform ran out of usable GPU memory above 1000 directions.}
    \label{fig:performance_compare}
\end{figure*}

To validate the proposed implementation with a $\mathrm{O}(N)$ computational complexity and see if it sufficiently performs for embedded or edge-computing devices, execution times were measured across a range of NVIDIA GPU platforms. We tested 5 different NVIDIA Jetson systems, which are embedded computing boards that can be considered relatively low-power devices (\qtyrange[range-units=single,range-phrase=-]{10}{60}{\W}). We also included two full size GPUs, with the RTX 3070 Ti M(obile) representing a high-end laptop GPU, and the RTX 3090 which represents a high-end desktop GPU. A single measurement of 163 840 samples (\SI{0.0364}{\s} at \SI{450}{\kHz}) was used and processed for 2, 10, 100, 200, 500, 1000, 2000, 3000 and 4000 directions of interest, each of which were executed 50 times for statistical comparison. The average timing results for 1000 directions can be found in Table \ref{tab:perf_all_dmas}, where we also show the relative increase when increasing the order as a percentage. In Fig. \ref{fig:performance_compare}a these same results are shown graphically, specifically for the middle-range GPU platform Jetson Orin Nano, visualizing the differences between the beamforming algorithms. Furthermore, Fig. \ref{fig:performance_compare}b shows the average execution times across all platforms, specifically for DMAS5-CF. These results clearly show the linear correlation between execution time and amount of directions. Table \ref{tab:perf_cf_vs_nocf} highlights the increase in average execution time for 1000 directions when applying the coherence factor, showing a consistent average increase of \SI{1.6}{\percent} in execution time. Finally, these timing results, together with the visual comparison of the beamforming methods shown earlier indicate that for a given platform, real-time performance is possible on the embedded platforms. 

\begin{table}
\caption{Beamforming - Average Execution Times for 1000 Directions}
\label{tab:perf_all_dmas}
\resizebox{\columnwidth}{!}{
\begin{tabular}{l|r|r|r|r|r|r|r|r|r}
\textbf{GPU Type} & \textbf{DAS} & \multicolumn{2}{c|}{\textbf{DMAS2}} & \multicolumn{2}{c|}{\textbf{DMAS3}} & \multicolumn{2}{c|}{\textbf{DMAS4}} & \multicolumn{2}{c}{\textbf{DMAS5}} \\ \cline{3-10}
  & \textbf{ms} & \textbf{ms} & $\mathbf+\Delta\%$ & \textbf{ms} & $\mathbf+\Delta\%$ & \textbf{ms} & $\mathbf+\Delta\%$ & \textbf{ms} & $\mathbf+\Delta\%$ \\ \hline
\textbf{Jetson Nano} & 350 & 661 & 89 & 1586 & 140 & 2715 & 72 & 4123 & 52 \\
\textbf{Jetson TX2NX} & 182 & 270 & 49 & 595 & 121 & 995 & 68 & 1500 & 51 \\
\textbf{Jetson Xavier NX} & 55 & 110 & 100 & 274 & 150 & 464 & 70 & 676 & 46 \\
\textbf{Jetson Orin Nano} & 66 & 124 & 89 & 288 & 132 & 488 & 70 & 719 & 48 \\
\textbf{Jetson Orin AGX} & 20 & 37 & 84 & 79 & 116 & 129 & 63 & 184 & 42 \\
\textbf{RTX 3070 Ti M} & 18 & 24 & 33 & 45 & 86 & 65 & 47 & 83 & 28 \\
\textbf{RTX 3090} & 20 & 21 & 7 & 26 & 24 & 32 & 27 & 40 & 24 \\
\end{tabular}%
}
\end{table}

\begin{table}[]
\caption{Coherence Factor - Average Execution Times for 1000 Directions}
\label{tab:perf_cf_vs_nocf}
\resizebox{\columnwidth}{!}{
\begin{tabular}{l|r|r|r}
\multicolumn{1}{r|}{\textbf{GPU Type}} & \textbf{DMAS (ms)} & \textbf{DMAS-CF (ms)} & \textbf{+$\Delta\%$ by CF} \\ \hline
\textbf{Jetson Nano}                   & 661                & 666                   & 0.7\%                      \\
\textbf{Jetson TX2NX}                  & 270                & 274                   & 1.7\%                      \\
\textbf{Jetson Xavier NX}              & 110                & 112                   & 1.9\%                      \\
\textbf{Jetson Orin Nano}              & 124                & 126                   & 1.7\%                      \\
\textbf{Jetson Orin AGX}               & 37                 & 38                    & 2.2\%                      \\
\textbf{RTX 3070 Ti M}                 & 24                 & 25                    & 2.5\%                      \\
\textbf{RTX 3090}                      & 21                 & 21                    & 0.4\%                     
\end{tabular}%
}
\end{table}

\section{Conclusion}
\label{sec:conclusion}
In this paper, we investigated the applicability of the non-linear Delay-Multiply-And-Sum (DMAS) beamformers to in-air acoustic imaging, given their successful adoption in medical imaging. We first introduced the mathematical formulation of these non-linear beamforming methods, which provide a more general alternative to the classical super-resolution techniques in broadband, single-snapshot scenarios that are encountered in this application domain. Indeed, the more commonly used MVDR or MUSIC have a strong dependence on the availability of multiple snapshots for estimating the signal covariance matrix, which is not possible in in-air acoustic imaging due to the slow speed of sound in air. Alternatives using spatial smoothing exist, but place strict constraints on the array geometry. Finally, sparsity-based techniques require an iterative reconstruction step, which is not feasible in real-time robotic applications.

While the proposed DMAS-CF method offers superior contrast, it is important to acknowledge the trade-offs inherent to non-linear multiplicative beamforming. These methods are known to be sensitive to off-axis clutter, which can produce artifacts due to non-linear cross-terms. In the context of in-air acoustic imaging, the impact of these artifacts is generally less pronounced because the imaging scenes are typically sparse—comprising a limited number of discrete sources or reflectors rather than the dense, continuous scattering as seen, for example, in the biomedical field when imaging biological tissue. Furthermore, because artifacts resulting from the multiplication of uncorrelated off-axis clutter lack spatial coherence across the array, the proposed coherence factor acts as a spatial filter that effectively suppresses these non-linear artifacts while preserving the main-lobe integrity. The simulated and real-world results conclusively demonstrate that non-linear beamforming is a powerful tool for high-quality in-air acoustic imaging. 

\newpage

By incorporating the DMAS principle, and further extending it through higher-order formulations, we successfully overcome the primary limitations of conventional beamforming, which is the limited dynamic range in the image due to the relatively high sidelobe energy. In-air acoustic sensing has a high dynamic range, which requires a high dynamic range in the imaging function to yield acoustic images that are useful in practical applications. In addition, the coherence factor post-processing step additionally improves the image contrast when applied to the beamformer outputs. The key contribution of this non-linear approach is the significant suppression of incoherent signals and sidelobe artifacts, leading to a substantial enhancement in image contrast across real-world, broadband measurements. 

Finally, our efficient GPU implementation, made possible by leveraging the Newton-Girard identities, ensures that this superior image quality is achieved without sacrificing the real-time performance, required for practical 3D sonar applications on embedded platforms in robotic settings. Through direct CUDA implementations, we have demonstrated the applicability of these algorithms in real-world sonar sensors based on an NVIDIA Jetson embedded GPU platform.

In future work, we aim to further explore the larger design space of acoustic imaging systems, taking into account different array geometries to quantify the mutual influence between layout, beamformer operation, and resulting image parameters. Furthermore, we wish to validate DMAS methods against traditional techniques like MVDR and MUSIC. However, these established methods generally require multiple snapshots and narrowband models, making them less applicable to in-air sonar sensing where large signal bandwidths and sparse topologies are beneficial for practical imaging.

\newpage
\bibliographystyle{IEEEtran}
\bibliography{main}
\end{document}